\documentclass[multphys,vecphys]{svmult}

\usepackage{makeidx}     
\usepackage{graphicx}    
\usepackage{multicol}    

\makeindex             

\newfont{\rmsmall}{cmr10 scaled 900}
\def\deg{\hbox{$^\circ$}}

\def\la{\mathrel{\hbox{\rlap{\hbox{\lower4pt\hbox{$\sim$}}}\hbox{$<$}}}}
\def\ga{\mathrel{\hbox{\rlap{\hbox{\lower4pt\hbox{$\sim$}}}\hbox{$>$}}}}

\def\fdg{\hbox{$.\!\!^\circ$}}
\def\farcm{\hbox{$.\mkern-4mu^\prime$}}

\def\apj{{ApJ}}                 
\def\apjl{{ApJ}}                
\def\apjs{{ApJS}}

\def\mnras{{MNRAS}}

\def\pasp{{PASP}}

\begin{document}

\title*{Results from the AST/RO Survey of the Galactic Center Region}
\titlerunning{AST/RO Galactic Center Survey}
\author{Antony A. Stark\inst{1},
Christopher L. Martin\inst{1},
Wilfred~M.~Walsh\inst{1},
Kecheng~Xiao\inst{1},
Adair~P.~Lane\inst{1},
Christopher~K.~Walker\inst{2}\and
J\"{u}rgen~Stutzki\inst{3}}
\authorrunning{Stark et al.}
\institute{Smithsonian Astrophysical Observatory, 60 Garden St. MS 12, Cambridge, MA 02138 USA
\texttt{aas, cmartin, wwalsh, kxiao, alane@cfa.harvard.edu}
\and Steward Observatory, University of Arizona, Tucson, AZ 85721 USA \texttt{cwalker@as.arizona.edu}
\and Physikalische Institut I, Universit\"{a}t zu K\"{o}ln, Z\"{u}lpicher
Stra{\ss}e 77, K\"{o}ln, D-50937 Germany \texttt{stutzki@ph1.uni-koeln.de}
}
%
%
\maketitle

We have used the 
Antarctic Submillimeter Telescope and Remote Observatory (AST/RO),
a 1.7m diameter single-dish submillimeter-wave telescope 
at the geographic South Pole, to determine the physical state
of gas in the Galactic Center region and assess its stability.
\index{AST/RO}
\index{submillimeter}
We present an analysis based on data obtained as part of an
ongoing AST/RO key project: the large-scale mapping of the
dominant cooling lines of the molecular interstellar medium
in the Milky Way \cite{martin03,kim02,ojha01,mookerjea03}.
These data are released for general use.

\section{Introduction}
\label{sec:stark1}

The distribution of
molecular gas in the Galaxy is known from extensive
and on-going surveys in CO and $^{13}$CO $J=1\rightarrow0$ and
$J=2\rightarrow1$; these are
spectral lines which indicate the presence of molecular gas.
These lines alone do not, however, determine the
excitation temperature, density, or cooling
rate of that gas.
Observations of C\,{\rmsmall I}  and the mid-$J$ lines of CO and $^{13}$CO
provide the missing information, showing a more complete picture
of the thermodynamic state of the molecular gas, highlighting
the active regions, and looking into the dense cores.
AST/RO can measure
the dominant cooling lines of molecular material in
\index{cooling}
the interstellar medium:
the ${ {}^{\rm 3}\! P_{\rm 1}\rightarrow{}^{\rm 3}\! P_{\rm 0}}$
(492 GHz) and 
${ {}^{\rm 3}\! P_{\rm 2}\rightarrow{}^{\rm 3}\! P_{\rm 1}}$
(809 GHz)
fine-structure lines of atomic carbon (C\,{\rmsmall I})
\index{carbon}
and the $J=4\rightarrow3$ (461 GHz) and $J=7\rightarrow6$ (807 GHz) rotational lines 
of carbon monoxide (CO).
These measurements can then be modeled
using the large velocity gradient
(LVG) approximation, and the gas temperature and density thereby determined.
Since the low-$J$ states of CO are in local thermodynamic
equilibrium (LTE) in almost all molecular gas,
measurements of mid-$J$ states are critical to
achieving a model solution
of the radiative transfer
by breaking the degeneracy between beam filling factor
and excitation temperature.

\section{AST/RO Survey Parameters}
\label{sec:stark2}

AST/RO
has been operational in the submillimeter-wave atmospheric windows
since 1995 \cite{stark01,stark03a}.
Essential to AST/RO's capabilities is 
its location at
Amundsen-Scott South Pole Station, an exceptionally cold,
dry site
which has unique logistical opportunities and challenges.
Among the key AST/RO projects is mapping of 
the Galactic Center Region.
\index{galactic center}
Sky coverage as of 2002 is
$-1 \fdg 3 < \ell < 2 \deg$,
$-0 \, \fdg 3 < b < 0 \, \fdg 2$ with
$0 \farcm 5$ spacing, resulting in spectra of three transitions at
24,000 positions on the sky  
\cite{kim02,martin03}.
The data are available 
on the AST/RO website\footnote{{\tt http://cfa-www.harvard.edu/ASTRO}}
for general use.

\begin{figure}[t!]
\centering
\includegraphics[width=10.8cm]{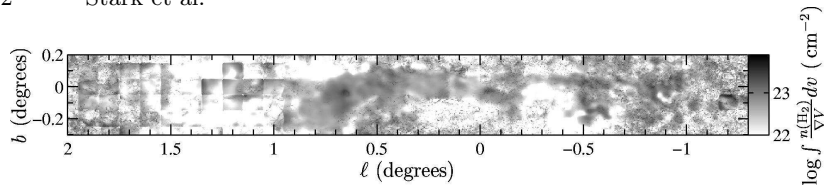}
\caption{Greyscale representation of molecular column density in the Galactic Center
Region, from an LVG model using AST/RO survey data \cite{martin03}.}
\label{fig:1}  
\end{figure}
\begin{figure}[t!]
\centering
\includegraphics[width=11.5cm]{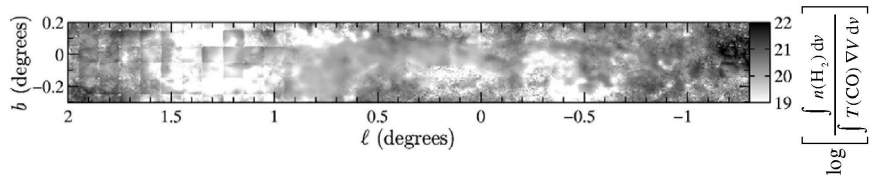}
\caption{Greyscale representation of the ratio of molecular column density
(as derived from an LVG model, cf. Fig. 1) to the integrated
brightness
of the CO $J = 1 \rightarrow 0$ line \cite{bally87b}.
If both these quantities were accurate measures of molecular column
density, then the ratio would be a constant; the ratio actually
varies by an order of magnitude either way around the mean.
}
\label{fig:2} 
\end{figure}

The $\mathrm{C}\,\scriptstyle{\rm I}$ emission has a spatial extent 
similar to the low-$J$ CO emission, but is more 
diffuse \cite{ojha01,mookerjea03}.
The CO $J = 4 \rightarrow 3$
emission is also found to be essentially coextensive with
lower-$J$ transitions of CO, indicating that even the $J=4$ state is
in LTE at most places; in contrast, the CO $J = 7 \rightarrow 6$ emission
is spatially confined to far smaller regions \cite{kim02}.
Applying an LVG model to these data \cite{martin03} together with
\index{LVG}
data from the Bell Labs 7m \cite{bally87b,bally88a}
yields maps
of gas density and temperature as a
function of position and velocity for the entire region.
\index{pressure}
\index{temperature}
Kinetic temperature is found to decrease from
relatively high values ($>70$ K) at cloud edges to lower values ($<50$ K)
in the interiors.
Typical pressures in the Galactic Center gas are
$n(\mathrm{H_2}) \cdot T_{kinetic} \sim 10^{5.2} \, \mathrm{K \, cm^{-3}}$.

An estimate of the molecular column density
\index{column density}
\index{density}
can be
obtained by integrating the LVG-derived density
over all velocities \cite{martin03}, as shown in Figure~\ref{fig:1}. 
An alternative estimate can be made utilizing the common assumption
that the column
density is proportional to the brightness of the
$J = 1 \rightarrow 0$ CO line \cite{liszt84,sanders84}.
Both of these estimates can be carried out on each line
of sight, and their ratio plotted as a function of
position, as shown in Figure \ref{fig:2}.
If the methodologies were mutually consistent, then
the ratio should not vary with position.  In
fact the ratio varies 
{\em{by two orders of magnitude}}.
Both methods incorporate rather ill-determined
multiplicative constants, which can arbitrarily
be adjusted to bring the two methods into agreement over some 
of the map but not all of it.
The discrepancies are caused by
variations in the excitation and optical depth of the
CO lines, suggesting that the LVG method should be
the more accurate method when submillimeter data are available.

\section{Galactic Center Gas Stability}
\label{sec:stark3}

The dynamics of the Galactic Center region depend critically
on whether the interstellar medium is self-gravitating or not
\cite{jenkins94};
given our estimate for the density of the molecular gas, we
can now determine its stability against coagulation into
self-gravitating lumps.
Binney et al. \cite{binney91} have suggested that gas in
the Galactic Center will be found mostly on closed orbits,
because gas thrown into a stellar system will tend to
shock and violently damp until it settles onto closed,
parallel streamlines.
In the central bar of the Milky Way, the closed orbits fall into
two families:
the elongated $x_1$ orbits running parallel to the outer parts
of the bar about 1 or 2 kpc from
the center, and the roughly circular $x_2$ orbits located near
the inner Lindblad resonance (ILR) of the bar at $\sim 350 \,
\index{Lindblad resonance}
{\mathrm{pc}}$  
and further inwards \cite{bissantz03}.
Gas falling into the system from outside first settles
onto the  $x_1$ orbits.  As it continues to lose energy, it 
slides downwards through $x_1$ orbits of
decreasing size until it gets to the inner $x_1$ orbits,
which are self-intersecting, where it transitions onto the
outer $x_2$ orbits \cite{bissantz03,jenkins94}
near the ILR---there it will tend to
accumulate, because inwards of the ILR
the net torque on the gas by the bar reverses sign, so that
gas inside the ILR is pushed outward while
gas outside the ILR is pushed inward \cite{elmegreen94}.
Gas therefore accumulates on the 
outer $x_2$ orbits near the ILR, growing more and more dense
until it becomes self-gravitating at a density 
\cite{elmegreen94} of
$$n({\mathrm{H_2}}) > {{0.3 \kappa^2}\over{m_p G}}
\approx 10^{3.4} \, \mathrm{cm ^{-3}}\, ,$$
where $\kappa \approx 900 \, {\mathrm{km \, s^{-1} \, kpc^{-1}}}$ is
the epicyclic frequency \cite{bissantz03}.
Gas on $x_2$ orbits appears in the data on a strip about
$50 \, {\mathrm{km \, s^{-1}}}$ wide extending from $(\ell, \, v) =
( -1 \deg , \, -80 \, {\mathrm{km \, s^{-1}}})$
to
$( +1 \deg , \, +80 \, {\mathrm{km \, s^{-1}}})$.
Our LVG model shows that the gas in this strip
is near the critical density, indicating that it
is only marginally stable
against gravitational coagulation into one or two 
giant clouds \cite{elmegreen94}.
\index{starbust}
This suggests a relaxation oscillator mechanism for starbursts
in the Milky Way, where
inflowing gas accumulates in a ring at 350 pc radius 
until the critical density is reached, and the resulting 
instability leads to the sudden deposition 
of a Giant Molecular Cloud onto the Galactic Center.

This work was supported in part by US NSF grant OPP-0126090.
We thank our colleagues R. A. Chamberlin, K. Jacobs, J. Kooi,
and G. A. Wright for their contributions to AST/RO.

%
%

\begin{thebibliography}{10}

\bibitem{martin03}
C.~L. Martin, W.~M. Walsh, K.~Xiao, A.~P. Lane, C.~K. Walker, and A.~A. Stark.
\newblock {\em \apjs}, in press 
\newblock {\tt astro-ph/0211025} (2004)

\bibitem{kim02}
S.~Kim, C.~L. Martin, A.~A. Stark, and A.~P. Lane.
\newblock {\em \apj}, \textbf{580}:896 (2002)

\bibitem{ojha01}
R.~Ojha, A.~A. Stark, H.~H. Hsieh, A.~P. Lane, R.~A. Chamberlin, T.~M. Bania,
  A.~D. Bolatto, J.~M. Jackson, and G.~A. Wright.
\newblock {\em \apj}, \textbf{548}:253 (2001)

\bibitem{mookerjea03}
B.~Mookerjea, C.~L. Martin, J.~Stutzki, A.~A. Stark, A.~P. Lane, W.~M. Walsh,
  and K.~Xiao.
\newblock In this volume. (2003)

\bibitem{stark01}
A.~A. Stark, J.~Bally, S.~P. Balm, T.~M. Bania, A.~D. Bolatto, R.~A.
  Chamberlin, G.~Engargiola, M.~Huang, J.~G. Ingalls, K.~Jacobs, J.~M. Jackson,
  J.~W. Kooi, A.~P. Lane, K.-Y. Lo, R.~D. Marks, C.~L. Martin, D.~Mumma,
  R.~Ojha, R.~Schieder, J.~Staguhn, J.~Stutzki, C.~K. Walker, R.~W. Wilson,
  G.~A. Wright, X.~Zhang, P.~Zimmermann, and R.~Zimmermann.
\newblock {\em \pasp}, \textbf{113}:567 (2001)

\bibitem{stark03a}
A.~A. Stark.
\newblock {AST}/{RO}: A small submillimeter telescope at the {S}outh {P}ole.
\newblock In T.~D. Oswalt, editor, {\em The Future of Small Telescopes in the
  New Millennium, Volume II---The Telescopes We Use}, pages 269--284. Kluwer
  Academic Publishers
\newblock {\tt astro-ph/0110429} (2003)

\bibitem{bally87b}
J.~Bally, A.~A. Stark, R.~W. Wilson, and C.~Henkel.
\newblock {G}alactic {C}enter molecular clouds. {I}. {S}patial and spatial
  velocity maps.
\newblock {\em \apjs}, \textbf{65}:13 (1987)

\bibitem{bally88a}
J.~Bally, A.~A. Stark, R.~W. Wilson, and C.~Henkel.
\newblock {\em \apj}, \textbf{324}:223 (1988)

\bibitem{liszt84}
H.~S. {Liszt}.
\newblock {\em Comments on Astrophysics}, \textbf{10}:137 (1984)

\bibitem{sanders84}
D.~B. {Sanders}, P.~M. {Solomon}, and N.~Z. {Scoville}.
\newblock {\em \apj}, \textbf{276}:182 (1984)

\bibitem{jenkins94}
A.~{Jenkins} and J.~{Binney}.
\newblock {\em \mnras}, \textbf{270}:703 (1994)

\bibitem{binney91}
J.~Binney, O.~E. Gerhard, A.~A. Stark, J.~Bally, and K.~I. Uchida.
\newblock {\em \mnras}, \textbf{252}:210 (1991)

\bibitem{bissantz03}
N.~{Bissantz}, P.~{Englmaier}, and O.~{Gerhard}.
\newblock {\em \mnras}, \textbf{340}:949 (2003)

\bibitem{elmegreen94}
B.~G. {Elmegreen}.
\newblock {\em \apjl}, \textbf{425}:L73 (1994)

\end{thebibliography}

\printindex
\end{document}